\documentclass[journal]{IEEEtran}

\usepackage{xr-hyper} 
\usepackage{hyperref}

\usepackage[acronym,nonumberlist,nogroupskip,style=super]{glossaries}
\usepackage{hyperref}
\newacronym{iot}{IoT}{Internet of Things} 

\usepackage{cite}
\usepackage{amsmath,amssymb,amsfonts}
\usepackage{graphicx}
\usepackage{textcomp}
\usepackage{bmpsize}
\usepackage[dvipsnames]{xcolor}
\usepackage{lipsum}

\usepackage[font=footnotesize,labelfont=bf]{caption}

\usepackage[export]{adjustbox}
 
\usepackage{pifont}
\usepackage{multirow}
\usepackage{tabularx}

\usepackage{algorithmicx}
\usepackage[noend]{algpseudocode}
\usepackage[T1]{fontenc}
\usepackage{algorithm} 
\usepackage{subfig}	
\usepackage{soul}
\usepackage{booktabs}

\usepackage{multirow}
\usepackage{rotating}
\usepackage{booktabs}

\usepackage{siunitx,etoolbox}

\usepackage{color, colortbl}

\usepackage{comment}

\begin{document}

\title{IoT Device Identification with Machine Learning: Common Pitfalls and Best Practices}

\author{ Kahraman Kostas, Rabia Yasa Kostas}

\maketitle

\begin{abstract}
This paper critically examines the device identification process using machine learning, addressing common pitfalls in existing literature. We analyze the trade-offs between identification methods (unique vs. class-based), data heterogeneity, feature extraction challenges, and evaluation metrics. By highlighting specific errors—such as improper data augmentation and misleading session identifiers—we provide a robust guideline for researchers to enhance the reproducibility and generalizability of IoT security models.
\end{abstract}

\begin{IEEEkeywords}
IoT security, IoT fingerprinting, machine learning, device identification
\end{IEEEkeywords}

\section{Introduction}

IoT devices bridge the cyber and physical worlds, enabling remote management and automation~\cite{al2015internet, patel2018internet}. However, their rapid proliferation and heterogeneous nature---varying significantly in hardware, OS, and interfaces---pose severe security challenges that traditional computing solutions cannot address. Lacking standard interfaces and possessing limited resources (battery, processor), these devices often prevent users from implementing necessary security measures manually. Consequently, \textit{IoT device identification} has emerged as a crucial method to infer device identity (e.g., brand and model) from network behavior, enabling tailored security policies, automated updates, and isolation strategies.

\begin{figure}[ht]
	\centerline{\includegraphics[width=1\columnwidth]{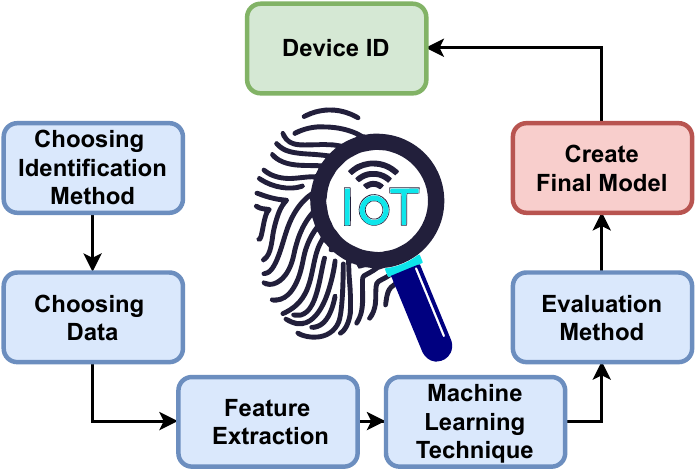}}
	\caption{Steps of device identification process.}
	\label{fig:stepsofID}
\end{figure}

In this study, we deconstruct the device identification process into four critical steps (see Figure~\ref{fig:stepsofID}): Method Selection, Data Preparation, Feature Extraction, and Evaluation. We aim to highlight prevalent mistakes made in each stage and propose mitigation strategies to ensure model robustness.

The remainder of this paper details identification taxonomies (Section~\ref{section:Methods}), data integrity considerations (Section~\ref{section:data}), feature extraction key points (Section~\ref{section:Features}), machine learning technique selection (Section~\ref{section:ML}), and evaluation metrics (Section~\ref{section:Evaluation}).

\section{Familiarizing with Methods}\label{section:Methods}

This section outlines the critical decisions regarding the scope of the study and the granularity of identification, highlighting how these choices dictate feature selection and model generalizability.

\subsection{Defining the Scope}
The definition of IoT varies significantly across the literature, ranging from low-power Wireless Sensor Networks (WSN) to complex multimedia devices~\cite{hussain2020machine}. A common pitfall in identification studies is failing to explicitly define these boundaries. Researchers must clearly state whether their dataset includes traditional IT assets (e.g., routers, smartphones) or is strictly limited to embedded IoT appliances. This decision directly impacts the heterogeneity of the traffic and the applicability of the proposed model to real-world scenarios.

\subsection{Identification Approaches and Trade-offs}
\begin{figure}[ht]
	\centerline{\includegraphics[width=1\columnwidth]{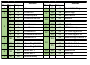}}
	\caption{Labelling the Aalto dataset according to 3 identification approaches\cite{kostas2024behaviour}.}
	\label{fig:comparedata}
\end{figure}

Determining the classification granularity is a pivotal step that affects both the labeling strategy and the required feature sets. As illustrated in Figure~\ref{fig:comparedata} using the Aalto University dataset~\cite{aalto2017dataset}, the same physical setup can yield vastly different class counts (from 15 to 33) depending on the chosen perspective. We categorize these approaches into three levels~\cite{yadav2020position}:

\begin{itemize}
	\item \textbf{Unique Identification:} Treats every single device as a distinct class, even identical models~\cite{hamad2019iot}.
	\textit{Trade-off:} This approach typically requires \textbf{flow-based features} to capture session-specific behaviors, as packet headers alone may not carry enough entropy to distinguish identical hardware. Consequently, this reduces the model's generalizability across different networks.
	
	\item \textbf{Type Identification:} Groups devices by make and model~\cite{miettinen2017iot}.
	\textit{Trade-off:} This allows for the use of \textbf{packet-based features}, offering higher generalizability. However, distinguishing devices with identical hardware but slightly different firmware versions remains a significant challenge.
	
	\item \textbf{Class Identification:} Aggregates functionally similar devices (e.g., cameras, plugs) under a single label~\cite{nguyen2019diot,aksoy2019automated}.
	\textit{Trade-off:} While this maximizes generalizability and allows for packet-based analysis, it introduces a dependency on \textit{expert knowledge} to manually define meaningful clusters based on hardware/software similarities.
\end{itemize}

\textbf{Recommendation:} Researchers should strictly align their feature extraction strategy with their identification goal. Attempting Unique Identification with only static packet headers will likely lead to overfitting, whereas using flow features for Class Identification may unnecessarily limit the model's scope.

\section{Familiarizing with Data}\label{section:data}

This section addresses the critical challenges in data acquisition, labeling integrity, and handling class imbalance to ensure reproducible results.

\subsection{Data Acquisition and Privacy}
Researchers generally rely on three sources: simulation, real-world testbeds, or public datasets. While simulations are cost-effective, they often fail to replicate the heterogeneous noise of real IoT environments. Conversely, collecting real-world data introduces severe privacy risks. Researchers must prioritize ethical considerations, anonymizing sensitive identifiers like IP addresses and user IDs. Alternatively, utilizing established benchmarks (e.g., UNSW IoT traces or datasets from Kaggle/UCI) facilitates comparability with existing literature. regardless of the source, the dataset must primarily consist of benign traffic to establish baseline behavior, though "normal" segments from attack datasets can be repurposed with caution.

\subsection{Labeling Integrity and Non-IP Devices}
A fundamental prerequisite is accurate ground-truth labeling. A common pitfall is relying on MAC or IP addresses as proxy labels for device identity. This is particularly misleading in heterogeneous networks involving non-IP protocols (e.g., ZigBee, Z-Wave).

\textbf{Case Study:} In the Aalto University dataset, a \textit{HueSwitch} connects via ZigBee to a \textit{HueBridge}. The bridge encapsulates the traffic and forwards it over Ethernet. Consequently, packets from the Switch carry the Bridge's MAC address. We identify this loss of distinct identity at the gateway level as the \textbf{Transfer Problem}~\cite{IoTDevID,kostas2023externally}. Relying solely on MAC addresses in such scenarios would erroneously label distinct devices as a single entity. Therefore, labels must be derived from the logical device identity, not just network headers.

\subsection{Handling Imbalance and Data Leakage}
\begin{figure}[ht]
	\centerline{\includegraphics[width=1\columnwidth]{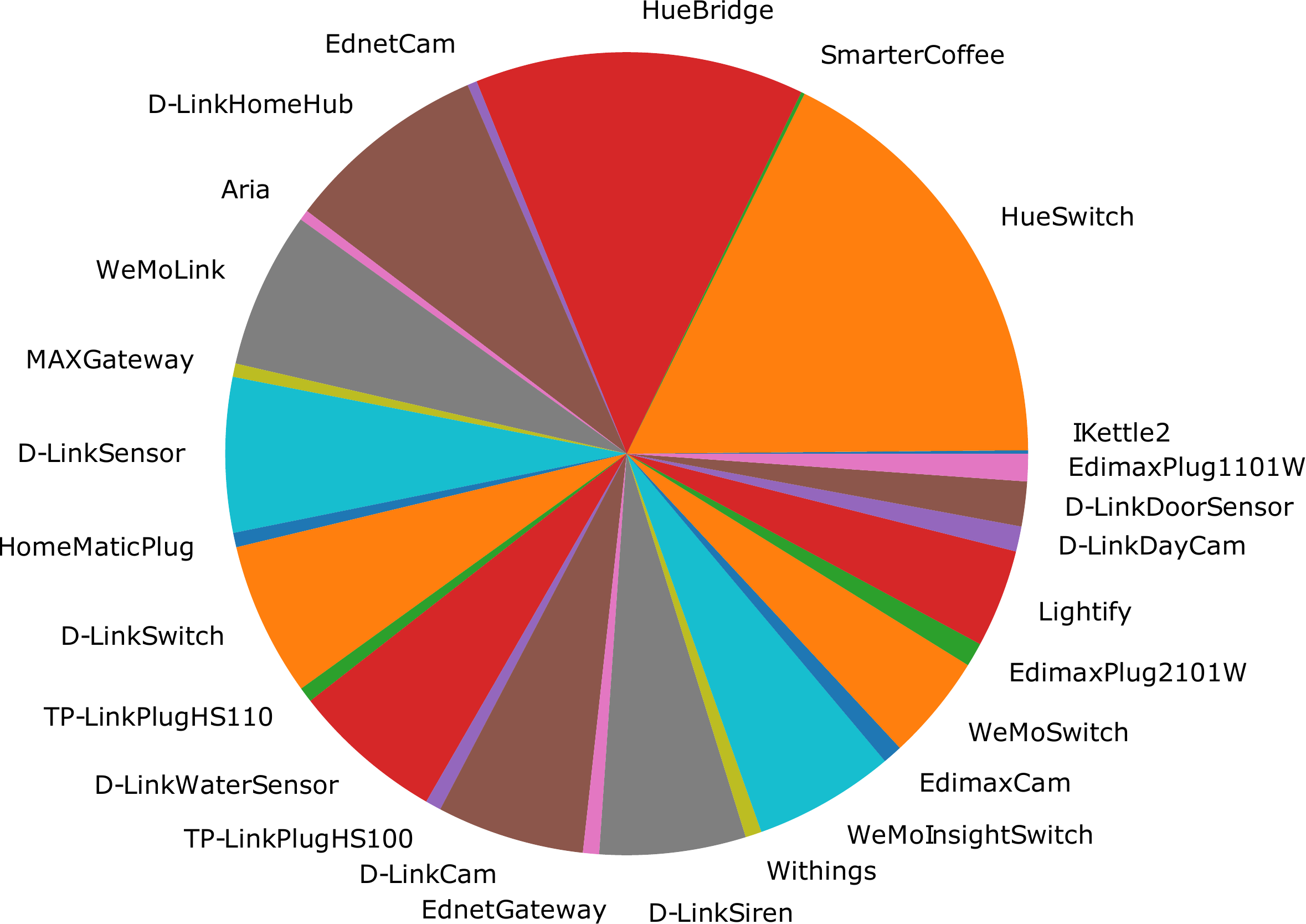}}
	\caption{The number of packets produced by the devices in the Aalto dataset, illustrating severe class imbalance.}
	\label{fig:aalto}
\end{figure}

IoT networks exhibit extreme class imbalance; a camera may generate GBs of data while a sensor transmits only bytes (see Fig.~\ref{fig:aalto}). While data augmentation is a valid strategy to mitigate this, it introduces a high risk of \textbf{data leakage}.

\textbf{Critical Rule:} The split between Training and Test sets must occur \textit{before} any augmentation. Augmenting the entire dataset and then splitting causes synthetic variations of the same sample to appear in both sets, inflating performance metrics artificially. Furthermore, the test set must remain pristine; resampling or augmenting test data destroys its representation of real-world distributions.

\section{Familiarizing with Feature Extraction}\label{section:Features}

Feature extraction is the bridge between raw data and the ML model. This section highlights efficiency strategies and critical feature selection pitfalls that lead to overfitting.

\subsection{Efficiency in Preliminary Analysis}
Processing gigabytes of PCAP data requires efficient tooling. While Python-based libraries like \href{https://dpkt.readthedocs.io/en/latest/}{dpkt} and \href{https://scapy.net/}{Scapy} offer ease of use and documentation, they often suffer from significant performance bottlenecks on large datasets. For preliminary analysis and filtering, we strongly recommend C-based command-line tools like \href{https://www.wireshark.org/docs/man-pages/tshark.html}{tshark}. Automated via bash scripts, tshark can perform statistical analysis and field extraction orders of magnitude faster than interpreted languages.

\subsection{Eliminating Identifying Features (Overfitting Risks)}
A robust model must learn device \textit{behavior}, not static identity. Including identifiers leads to "shortcut learning," where the model memorizes the device rather than classifying its type. Researchers must rigorously prune the following:

\begin{itemize}
	\item \textbf{Explicit Identifiers:} MAC and IP addresses must be removed. They identify a specific physical unit or network location, offering zero generalizable information about the device type.
	
	\item \textbf{Session Identifiers:} Ephemeral values such as \textbf{Source Ports, TCP Sequence Numbers, and IP IDs} are generated randomly per session. A model trained on these will fail on new sessions. If session-specific features are unavoidable, strict separation of sessions between training and testing sets is required to prevent data leakage.
	
	\item \textbf{Implicit Identifiers:} Often overlooked, \textbf{Header Checksums} must be discarded as they mathematically encode IP addresses and port numbers. Similarly, absolute timestamps can bias the model toward specific capture windows.
\end{itemize}

\subsection{Sanitizing Raw Data Inputs}
\begin{figure}[ht]
	\centerline{\includegraphics[width=1\columnwidth]{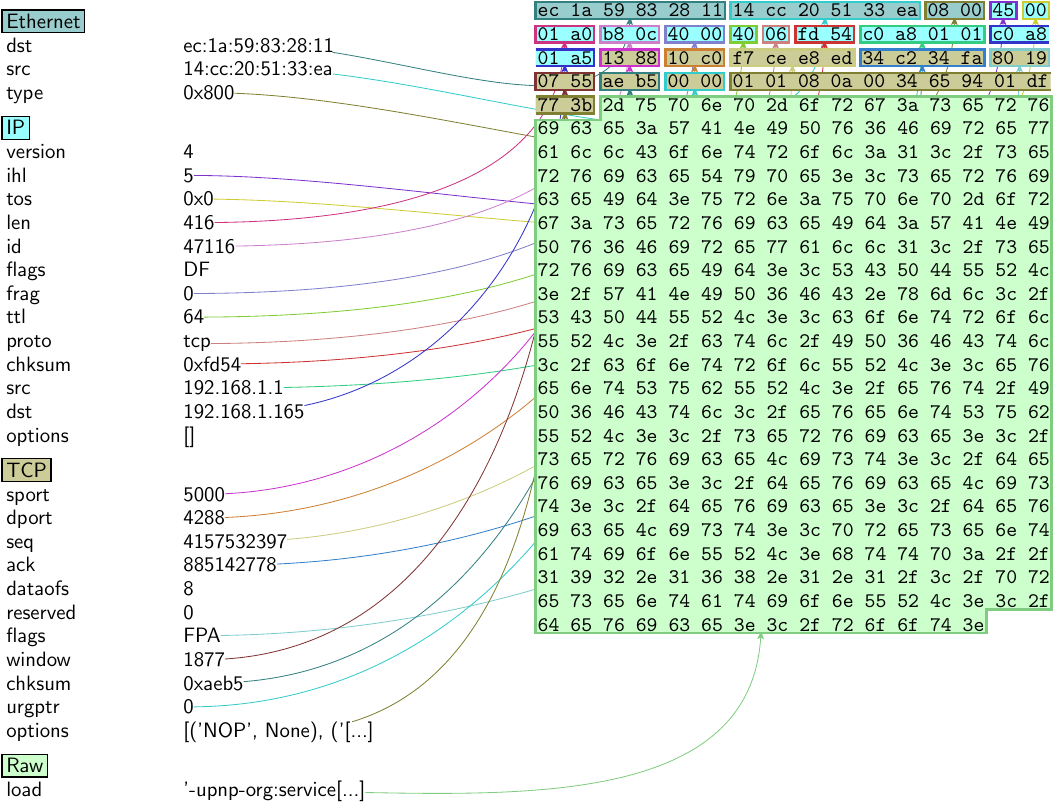}}
	\caption{The fields contained in a network packet and their byte equivalents.}
	\label{fig:bytes}
\end{figure}

Deep learning approaches often utilize raw packet bytes to minimize manual feature engineering. However, raw headers are not immune to the pitfalls mentioned above. Feeding a raw Ethernet/IP header to a neural network exposes it to the exact same MACs, IPs, and checksums. Therefore, when using raw data, researchers must implement a \textbf{masking/sanitization} step to zero out or strip these specific header fields before feeding them into the model.  A network packet converted to raw bytes is given in Fig.~\ref{fig:bytes}.

\section{Familiarizing with Machine Learning}\label{section:ML}

Selecting the appropriate machine learning architecture involves balancing accuracy against scalability, speed, and explainability. This section addresses architecture design and algorithm selection.

\subsection{Scalability and Model Architecture}
Device identification is inherently a multi-class problem. A monolithic multi-class model suffers from poor scalability; adding a single new device necessitates retraining the entire system. To mitigate this, we recommend a \textbf{One-vs-Rest (OvR)} approach (also known as One2All). By training a separate binary classifier for each device, the system becomes modular: new devices can be added by simply appending a new binary model without disturbing existing classifiers, significantly reducing maintenance overhead.

\subsection{Algorithm Selection and the "No Free Lunch" Theorem}
There is no universally superior algorithm~\cite{wolpert2002supervised}. Researchers should avoid the "Deep Learning defaultism." For tabular data lacking temporal dependencies—which characterizes most flow/packet feature sets—classical approaches like Decision Trees often outperform Deep Learning in both accuracy and training efficiency~\cite{lundberg2020local2global}. A robust strategy involves benchmarking representative algorithms from different families (e.g., Tree-based, Kernel-based, probabilistic) to identify the best fit for the specific data distribution.

\subsection{Operational Constraints: Speed and Interpretability}
\begin{figure}[ht]
	\centerline{\includegraphics[width=1\columnwidth]{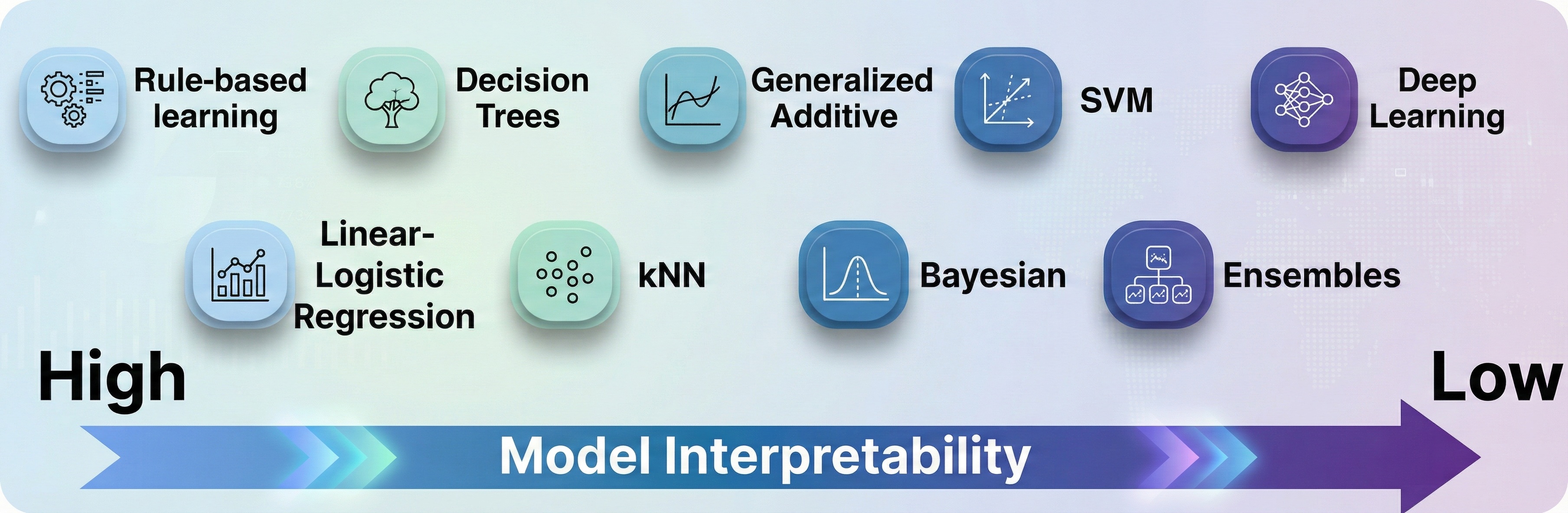}}
	\caption{Interpretability levels of common machine learning methods.}
	\label{fig:interpretability}
\end{figure}

In cybersecurity, high accuracy is insufficient if the model is too slow or opaque.
\begin{itemize}
	\item \textbf{Inference Time:} Network traffic operates at micro-second scales. Algorithms like k-NN or kernel-SVM, while accurate, often exhibit high inference latency (computational complexity grows with dataset size). Models must be evaluated on their real-time sorting capability, not just offline accuracy.
	\item \textbf{Interpretability:} Understanding \textit{why} a device was identified is crucial for security auditing. As shown in Figure~\ref{fig:interpretability}, there is often a trade-off between performance and interpretability. While Deep Learning offers high capacity, its "black-box" nature complicates feature analysis compared to transparent models like Decision Trees or Logistic Regression~\cite{arrieta2020explainable}.
\end{itemize}

\section{Familiarizing with Evaluation}\label{section:Evaluation}

The choice of evaluation metrics is not merely a preference but a statistical necessity dictated by the data distribution. This section addresses the pitfalls of standard metrics in IoT contexts.

\subsection{The Accuracy Trap in Imbalanced Datasets}
Accuracy is the most ubiquitous metric in literature, offering high comparability. However, in IoT datasets characterized by severe class imbalance (as discussed in Section~\ref{section:data}), accuracy is often misleading. In a scenario where one device constitutes 90\% of the traffic, a model that blindly predicts the majority class for every instance achieves 90\% accuracy while failing to identify any other device. Accuracy is a holistic metric; it masks the poor performance of minority classes (under-represented devices) behind the high volume of majority classes. Therefore, relying solely on accuracy is insufficient for robust device identification.

\subsection{Precision, Recall, and Standardization}
To overcome the limitations of accuracy, researchers should focus on granular and diagnostic metrics.

\begin{itemize}
	\item \textbf{Recall (Sensitivity):} This measures the ability to correctly identify a specific device. \textit{Note on Terminology:} Literature often refers to this as "detection rate," "recognition rate," or "identification rate." To avoid confusion, we recommend adhering to the standard ML terminology: \textbf{Per-class Recall}.
	
\item \textbf{The F1-Score and Macro-Averaging:} Using Recall alone is risky as it ignores False Positives. We recommend the \textbf{F1-Score} (harmonic mean of Precision and Recall) as the primary metric. Crucially, researchers should distinguish between averaging methods:
\begin{itemize}
	\item \textit{Avoid Micro-averaging:} It aggregates global counts of true/false positives, implicitly weighting the score by class frequency. In imbalanced sets, it mimics the bias of Accuracy, favoring majority classes.
	\item \textit{Use Macro-averaging:} It calculates the metric for each class independently and then averages them. This treats all classes equally regardless of size, ensuring that the poor performance of a minority device is exposed rather than hidden.
\end{itemize}
	\item \textbf{Visualizing Error Patterns (Confusion Matrix):} Scalar metrics cannot reveal \textit{why} a model fails. The \textbf{Confusion Matrix} is indispensable for diagnosing specific failure modes. It reveals inter-class misclassifications—identifying, for instance, if two devices from the same manufacturer are consistently confused due to shared firmware stacks—which global metrics might hide.
\end{itemize}
\section{Conclusion}
In this study, we critically examined the device identification pipeline, identifying systemic pitfalls that hinder the reproducibility and reliability of IoT security models. We highlighted that:
\begin{itemize}
	\item \textbf{Methodology:} The choice between Unique, Type, and Class identification must align with the feature set (flow vs. packet).
	\item \textbf{Data Integrity:} Privacy ethics and correct labeling (avoiding proxy labels like MAC addresses) are foundational.
	\item \textbf{Feature Engineering:} Static identifiers (IP, MAC, Session IDs) must be rigorously pruned to prevent overfitting.
	\item \textbf{Evaluation:} In imbalanced IoT environments, F1-Score provides a more honest assessment than Accuracy.
\end{itemize}
By addressing these common mistakes, researchers can build more generalized, scalable, and practical security solutions for the expanding IoT landscape.

\bibliographystyle{IEEEtran}
\bibliography{references}
\end{document}